# Interplanetary Causes and Impacts of the 2024 May Superstorm on the Geosphere: An Overview

Rajkumar Hajra[1][2], Bruce Tsatnam Tsurutani[3], Gurbax Singh Lakhina[4], Quanming Lu[2], Aimin Du[5]

## Abstract

The recent superstorm of 2024 May 10–11 is the second largest geomagnetic storm in the space age and the only one that has simultaneous interplanetary data (there were no interplanetary data for the 1989 March storm). The May superstorm was characterized by a sudden impulse (SI$^+$) amplitude of +88 nT, followed by a three-step storm main phase development which had a total duration of ~9 hr. The cause of the first storm main phase with a peak SYM-H intensity of -183 nT was a fast forward interplanetary shock (magnetosonic Mach number $M_{ms}$ ~7.2) and an interplanetary sheath with southward interplanetary magnetic field component $B_s$ of ~40 nT. The cause of the second storm main phase with a SYM-H intensity of -354 nT was a deepening of the sheath $B_s$ to ~43 nT. A magnetosonic wave ($M_{ms}$ ~0.6) compressed the sheath to a high magnetic field strength of ~71 nT. Intensified $B_s$ of ~48 nT was the cause of the third and most intense storm main phase with a SYM-H intensity of -518 nT. Three magnetic cloud events with $B_s$ fields of ~25–40 nT occurred in the storm recovery phase, lengthening the recovery to ~2.8 days. At geosynchronous orbit, ~76 keV to ~1.5 MeV electrons exhibited ~1–3 orders of magnitude flux decreases following the shock/sheath impingement onto the magnetosphere. The cosmic ray decreases at Dome C, Antarctica (effective vertical cutoff rigidity <0.01 GV) and Oulu, Finland (rigidity ~0.8 GV) were ~17% and ~11%, respectively relative to quite time values. Strong ionospheric current flows resulted in extreme geomagnetically induced currents of ~30–40 A in the sub-auroral region. The storm period is characterized by strong polar region field-aligned currents, with ~10 times intensification during the main phase, and equatorward expansion down to ~50° geomagnetic (altitude-adjusted) latitude.

[1] Corresponding author rajkumarhajra@yahoo.co.in, rhajra@ustc.edu.cn
[2] CAS Key Laboratory of Geospace Environment, School of Earth and Space Sciences, University of Science and Technology of China, Hefei, People's Republic of China
[3] Retired, Pasadena, California, USA
[4] Retired, B-701 Neel Sidhi Towers, Sector-12, Vashi, Navi Mumbai, India
[5] College of Earth and Planetary Sciences, Chinese Academy of Sciences, Beijing, People's Republic of China

## 1. Introduction

One of the main goals of this work is to identify the interplanetary causes and impacts of the 2024 May 10–11 geomagnetic storm that has attracted the attention of the science community and the general public. While display of auroras across Europe, Asia and America at low geomagnetic latitudes down to ~27.6° (Puerto Rico)[6] was of great interest to the public, the extremely high peak intensity of the storm makes it a rare event to study. More specifically, with a SYM-H peak intensity of -518 nT, it is the second strongest geomagnetic storm of the space age, and the only one that we have interplanetary plasma and magnetic field data for. The strongest storm is the 1989 March SYM-H = -720 nT storm, with no interplanetary measurements. We will show that the May storm was a unique three-step main phase storm and we identify the interplanetary causes of the three steps of the storm.

A geomagnetic storm and associated interplanetary events may initiate a chain of processes causing disturbances in the terrestrial magnetosphere, ionosphere and even on the ground (see Hajra et al. (2020) for some comprehensive case studies, and Tsurutani et al. (2020, 2023, 2024) for reviews of underlying physics). Thus, an integrated study of the solar, interplanetary, magnetospheric, ionospheric and ground-based observations is important for a comprehensive understanding of the sun-Earth coupled system. In this article, we will explore the rare and extreme superstorm of 2024 May and its impacts using near-Earth measurements of solar wind, radiation belt particles, ionospheric plasma along with ionospheric and ground-based current measurements. This study is aimed at enhancing our understanding of the causes, features and impacts of extreme space weather events on the Geosphere. Such a study is also important for developing predictive capability of such unique space weather events in future.

## 2. Data Analyses and Results

---

[6] See: https://spaceweather.com/archive.php?view=1&day=11&month=05&year=2024

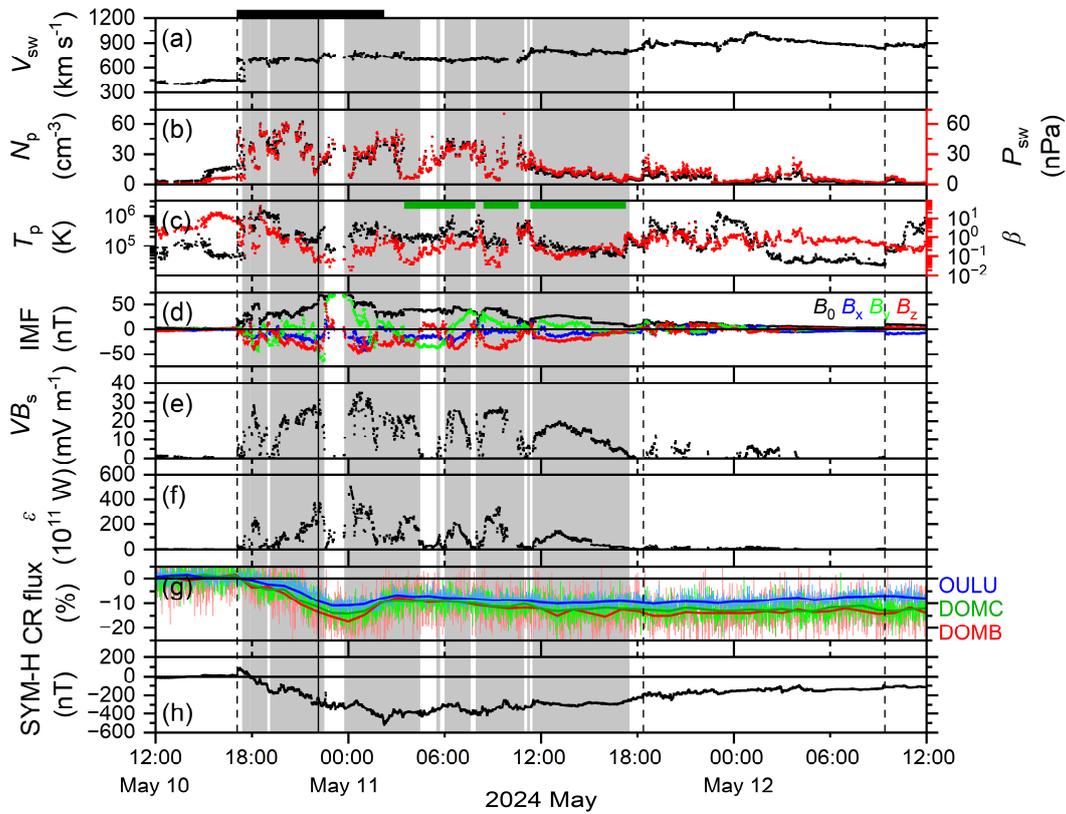

**Figure 1. Solar wind/interplanetary and geomagnetic conditions during the 2024 May storm.** From top to bottom, the panels show: (a) the solar wind plasma speed $V_{sw}$, (b) proton density $N_p$ (black, legend on the left) and ram pressure $P_{sw}$ (red, legend on the right), (c) proton temperature $T_p$ (black, legend on the left) and plasma-$\beta$ (red, legend on the right), (d) interplanetary magnetic field (IMF) magnitude $B_0$ and $B_x$, $B_y$, $B_z$ components, (e) electric field $VB_s$, (f) Akasofu $\varepsilon$-parameter, (g) the cosmic ray (CR) flux in percentage (normalized to the pre-storm values) for Dome C standard neutron monitor (DOMC), Dome C "bare" neutron monitor (DOMB) and Oulu neutron monitor (OULU), thin and bold curves represent 1-minute and 1-hr resolution data, respectively, and (h) geomagnetic SYM-H index during May 10–12. Vertical dashed lines indicate interplanetary fast forward shocks. A vertical solid line indicates a fast forward wave. The storm main phase is marked by a black horizontal bar at the top. IMF $B_s$ components are marked by light-gray vertical shadings. Magnetic clouds (MCs) with low β and smooth IMF $B_0$ are marked by green horizon bars in panel (c).

**2.1. Interplanetary Drivers**

The geomagnetic storm onset, development and recovery are studied using the temporal profile of the SYM-H index[7] (1-minute version of 1-hr Dst index; Sugiura 1964), whose decrease is considered to be a manifestation of storm-time enhancement of terrestrial equatorial westward ring current particle energy at ~2–7 Earth radius ($R_E$) (Sckopke 1966; Dessler & Parker 1969). From the geomagnetic SYM-H index variation (Figure 1h), the storm started with a sudden impulse (SI+; Araki et al. 1993; Tsurutani et al. 2011; Tsurutani & Lakhina 2014) of +88 nT at 17:15 UT on May 10. The SI+ is followed by a gradual SYM-H decrease indicating strong ring current growth (and the storm main phase development). The storm main phase is characterized by three major SYM-H peaks: -183 nT at 19:21 UT on May 10, -354 nT at 23:12 UT on May 10, and -518 nT at 02:14 UT on May 11. The third SYM-H peak is followed by the storm recovery. The recovery phase is characterized by multiple local SYM-H decreases, continuing up to ~end of May 13 (not shown). Thus, the durations of the storm main and recovery phases are ~9 hr and ~2.8 days, respectively.

The storm main phase onset is coincident with the onset of Forbush decreases (Forbush 1938) as observed in cosmic ray (CR) count rates[8] (Figure 1g) measured by the Dome C (Antarctica) standard neutron monitor (DOMC), the Dome C "bare" neutron monitor (DOMB) and the Oulu (Finland) neutron monitor (OULU). The Dome C (geomagnetic latitude: 88.8°S, longitude: 55.6°E, altitude above sea level: 3233 m) neutron monitors with very low effective vertical cutoff rigidity <0.01 GV registered peak CR decreases of ~14% (DOMC) and ~17% (DOMB), while the Oulu (geomagnetic latitude: 62.1°N, longitude: 115.9°E, altitude above sea level: 15 m) monitor with a cutoff rigidity ~0.8 GV registered a peak CR decrease of ~11%. The "classical" two-step decreases are prominent in the DOMB and DOMC data. The CR decreases are estimated from the average count rates on the pre-storm/geomagnetically quiet day May 9. The peak decrease corresponds roughly to the second SYM-H peak and the start of the third storm main phase. As usual, the CR decrease phase of ~7–12 hr is significantly faster than the slow and gradual recovery continuing for several weeks (e.g. Lockwood 1971).

Figures 1a–e show the near-Earth (at the Earth's bow shock nose) solar wind plasma and interplanetary magnetic field (IMF) variations[9]. All SYM-H decreases exhibit a one-to-one correlation with the IMF southward component $B_s$ (Figure 1d) and interplanetary motional (eastward) electric field $VB_s$ (Figure 1e). $B_s$ is defined as $-B_z$ for IMF $B_z$ component <0, and 0 for $B_z \geq 0$; $VB_s$ represents a motional electric field for $B_z$ <0; $V$ is plasma speed $V_{sw}$. $B_s$ components are marked by light-gray vertical shadings to show their correspondence with the SYM-H decreases. The three SYM-H peaks are characterized by $B_s$ peak values

---

[7] The 1-minute resolution SYM-H index data are collected from the World Data Center for Geomagnetism, Kyoto, Japan (https://wdc.kugi.kyoto-u.ac.jp/).
[8] The 1-minute and 1-hr resolution CR count rates are collected from the Cosmic Ray Station of the University of Oulu/Sodankyla Geophysical Observatory (https://cosmicrays.oulu.fi/).
[9] Near-Earth solar wind plasma and IMF measurements (1-minute resolution) are collected from NASA's OMNIWeb Plus (https://omniweb.gsfc.nasa.gov/ow_min.html).

($B_s$ component duration) of: 40.4 nT at 18:06 UT (1.6 hr), 43.4 nT at 22:12 UT (3.4 hr) on May 10; and 47.9 nT at 00:36 UT (4.7 hr) on May 11, respectively. The corresponding $VB_s$ peaks are: 28.7, 31.4 and 35.0 mV m$^{-1}$, respectively.

There are three strong and long-duration IMF $B_s$ intervals in the storm recovery phase, with peak Bs components of 38.7, 39.6, and 24.5 nT, respectively. Based on the low plasma-$\beta$ (the ratio of the plasma pressure to the magnetic pressure) of ~0.03–0.07, and high, smooth IMF $B_0$ (~28.5–41.6 nT) without discontinuities or waves, we postulate that these are parts of magnetic clouds (MCs; Burlaga et al. 1981; Klein & Burlaga 1982). They occurred from ~03:28 to 07:54 UT, from ~08:25 to 10:36 UT, and from ~11:19 to 17:17 UT on May 11. However although the solar wind energy input from these three events increased the energy of the Earth's ring current, they did not cause an increase in the storm peak intensity. They did however extend the length of the storm recovery phase.

Magnetic reconnection between IMF $B_s$ and northward geomagnetic fields at the Earth's dayside magnetopause (Dungey 1961) is considered to be the major mechanism for injection of solar wind energy to the terrestrial magnetosphere, that leads to the intensification of the terrestrial equatorial ring current, as depicted in the SYM-H decreases (Gonzalez et al. 1994). The first three intervals of intense $B_s$ created the three-step storm main phase. An empirical measure of magnetospheric energy input rate through magnetic reconnection is given by the Akasofu $\varepsilon$-parameter[10] (Perreault & Akasofu 1978). The three-step storm main phase is characterized by three major $\varepsilon$ peaks of ~2.5×10$^{13}$ W (at ~18:11 UT on May 10), ~3.7×10$^{13}$ W (at ~22:12 UT on May 10), and ~5.1×10$^{13}$ W (at ~00:08 UT on May 11) (Figure 1f).

**Table 1.** Characteristic features of the interplanetary discontinuities at the WIND spacecraft location.

| Date and Time[a] (UT) | Type[b] | SI[+] | Solar Wind and IMF Parameters Across the Discontinuity[c] | | | | | Discontinuity Parameters | |
|---|---|---|---|---|---|---|---|---|---|
| | | | $V_{sw}$ (km s$^{-1}$) | $N_p$ (cm$^{-3}$) | $P_{sw}$ (nPa) | $T_p$ (10$^4$ K) | $B_0$ (nT) | $V_{sh}$ (km s$^{-1}$) | $M_{ms}$ |
| May 10 16:37 (17:03) | FFS | 88 | 443–714 | 16.1–53.8 | 4.8–49.4 | 7.22–57.45 | 6.6–19.3 | 386 | 7.15 |
| May 10 21:40 (22:12) | FFW | 136 | 664–723 | 15.0–28.5 | 11.0–24.6 | 19.50–26.24 | 51.0–69.3 | 147 | 0.55 |
| May 11 18:02 (18:17) | FFS | 39 | 853–911 | 7.7–16.9 | 5.0–24.7 | 26.06–35.51 | 8.1–17.7 | 106 | 1.46 |
| May 12 09:08 (09:24) | FFS | 36 | 839–890 | 1.4–4.5 | 1.9–5.9 | 17.74–30.91 | 4.1–11.4 | 150 | 1.83 |

---

[10] $V_{sw}B_0^2 \sin^4(\theta/2)R_{CF}^2$, where θ is the IMF clock angle, $R_{CF}$ is magnetopause scale size (Chapman & Ferraro 1931), given by $R_E \left(\frac{2B_E^2}{\mu_0 m_p N_p V_{sw}^2}\right)^{\frac{1}{6}}$, where $B_E$ is the equatorial magnetic field on the Earth's surface, $\mu_0$ is the free-space permeability, and $m_p$ is the solar wind proton mass.

**Notes.**

[a] Times in the parentheses indicate identification times of the discontinuities at the Earth's bow shock nose, based on Figure 1.

[b] FFS is fast forward shock characterized by $M_{ms}$ >1, and FFW is fast forward wave characterized by $M_{ms}$ <1.

[c] The values corresponds to upstream to downstream of a discontinuity.

What are the sources of the IMF $B_s$? Analysis of the solar wind plasma and IMF parameters show four interplanetary discontinuities (marked by vertical dashed and solid lines in Figure 1) identified by simultaneous increases in solar wind plasma speed $V_{sw}$ (Figure 1a), proton density $N_p$ (Figure 1b), ram pressure $P_{sw}$ (Figure 1b), proton temperature $T_p$ (Figure 1c), and IMF magnitude $B_0$ (Figure 1d). The characteristic features of the interplanetary discontinuities, as identified at the location of the WIND spacecraft[11] upstream of the Earth at a distance of ~236–243 $R_E$, are listed in Table 1. The characteristic parameters are determined using the (plasma-IMF) mixed-mode discontinuity normal determination method (Abraham-Shrauner 1972) and the application of the Rankine–Hugoniot (Rankine 1870; Hugoniot 1887, 1889) conservation laws (detail description of the method can be found in Smith 1985; Tsurutani & Lin 1985; Tsurutani et al. 2011; Hajra et al. 2016, 2020, 2023; Hajra & Tsurutani 2018a; Hajra 2021). The discontinuity identified at ~17:03 UT (~16:37 UT at WIND) on May 10 is determined to be a fast forward shock (steepened magnetosonic wave), characterized by a magnetosonic Mach number $M_{ms}$ of ~7.2, moving at a shock speed $V_{sh}$ of ~386 km s$^{-1}$ relative to the upstream solar wind plasma. This shock caused the SI$^+$ observed in SYM-H at 17:15 UT. The discontinuity detected at ~22:12 UT on May 10 is found to be a fast wave, having an $M_{ms}$ of ~0.6, and moving at $V_{sh}$ of ~147 km s$^{-1}$. The wave is coincident with a sharp northward turning of IMF, leading to ring current particle loss, as observed in a sharp increase in SYM-H from -312 to -176 nT (SI$^+$ = 136 nT). Two more fast forward shocks are detected in the recovery phase, leading to local sharp increases in the SYM-H index from -227 to -188 nT (SI$^+$ = 39 nT), and from -130 to -94 nT (SI$^+$ = 36 nT). The shocks are presumably driven by ICMEs moving faster than the ambient upstream magnetosonic wave speed (Kennel et al. 1985; Tsurutani et al. 2011). The ICMEs are interplanetary counterparts of several coronal mass ejections (CMEs) erupted in association with X1.1–X5.8 class solar flares from the giant solar active region AR3664 on May 8–11. See Appendix A for further details on the active region and the flares. The shocks and the wave strongly compress the solar wind plasma and IMF, known as interplanetary sheaths (see Zurbuchen & Richardson 2006; Kilpua et al. 2017 for excellent reviews on ICMEs and associated near-Earth interplanetary structures). The observed IMF $B_s$ components are integral parts of the sheaths. The solar wind plasma and IMF parameters shown in Figure 1 indicate a shock-sheath and a wave-sheath interaction leading to strengthening of the $B_s$ components (peak $B_0$ ~71 nT) responsible for this giant geomagnetic storm. Thus, the shock-sheath and wave-sheath interactions are determined to be interplanetary causes of the 2024 May superstorm main phase.

---

[11] WIND measurements are obtained from NASA's Coordinated Data Analysis Web (CDAWeb: https://cdaweb.gsfc.nasa.gov/istp_public/).

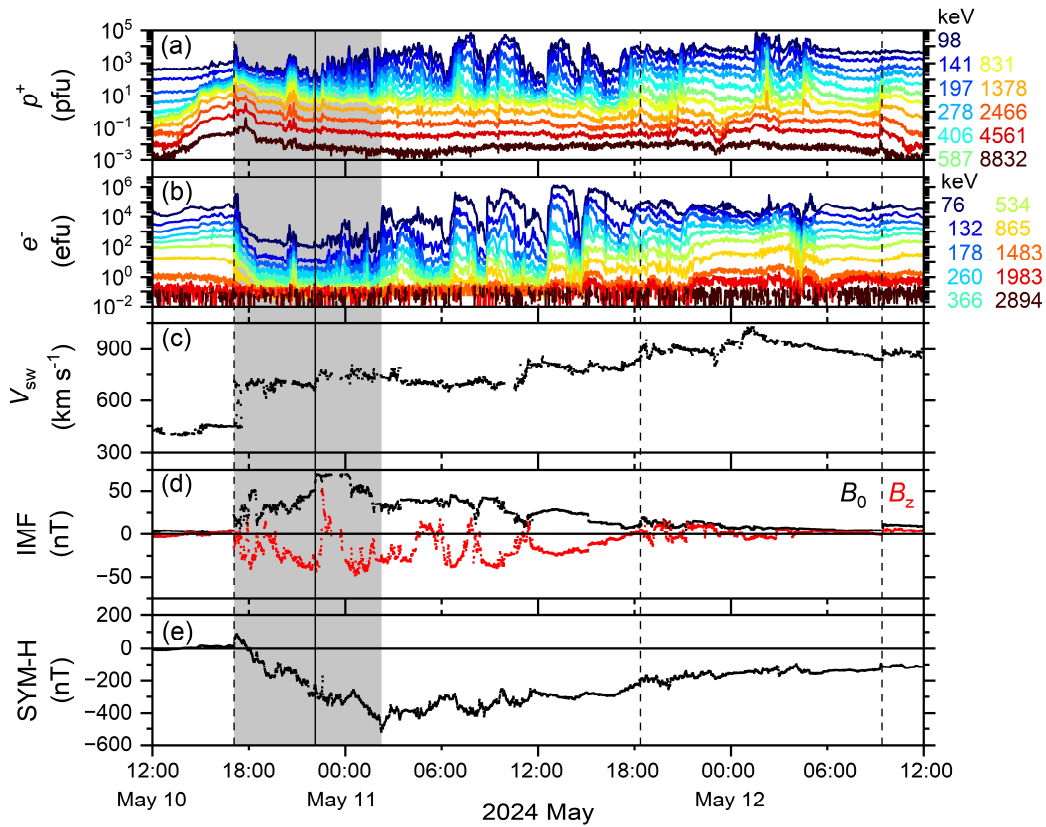

**Figure 2. Radiation belt evolution during the 2024 May storm.** From top to bottom, the panels show: (a) ~98 keV to ~8.8 MeV differential proton fluxes, and (b) ~76 keV to ~2.9 MeV differential electron fluxes at the geosynchronous orbit, (c) solar wind plasma $V_{sw}$, (d) IMF $B_0$ and $B_z$, (e) SYM-H index during May 10–12. The proton and electron energy values are marked by different colors as indicated on the right in panels (a) and (b). Panels (c)–(e) and markings of the shocks (vertical dashed lines) and the wave (vertical solid line) are repeated from Figure 1 for references. The storm main phase is marked by a light-gray shading.

**2.2. Radiation Belt Evolution**

The impacts of the storm and associated interplanetary events on the geosynchronous orbit energetic (~98 keV to ~8.8 MeV) proton and (~76 keV to ~2.9 MeV) electrons[12] are shown in Figures 2a and 2b, respectively. Following the fast forward shock at ~17:03 UT on May 10, the ~76–534 keV electrons exhibited ~3 orders of magnitude decreases in their fluxes, ~0.9 MeV electrons ~2 orders of magnitude decrease, and 1.5 MeV electrons ~1 order of magnitude decrease compared to their pre-shock fluxes (Figure 2b). No significant impact of the shock was recorded on the ~2.0–2.9 MeV electrons. The proton fluxes were more or less stable during the storm main phase, and the recovery phase is characterized by episodic injections of ~98–406 keV protons during southward IMF intervals (Figure 2a). The storm recovery phase is characterized by sporadic injections of ~76–534 keV electrons. The ~0.9–2.9 MeV electron fluxes exhibited slower and gradual increases in the storm recovery phase.

---
[12] Measured by GOES-18 (https://www.ngdc.noaa.gov/stp/satellite/goes-r.html).

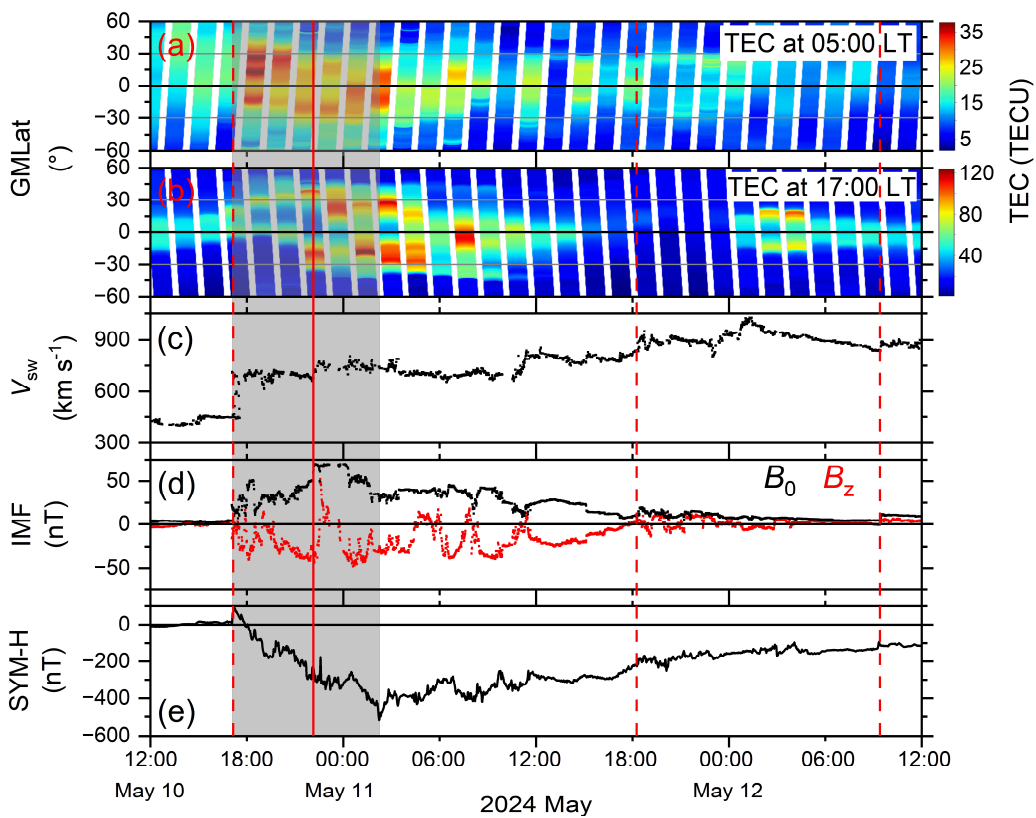

**Figure 3. Equatorial ionization anomaly during the 2024 May storm.** From top to bottom, the panels show: (a) variation of ionospheric total electron content (TEC) at 05:00 LT and (b) variation of TEC at 17:00 LT with geomagnetic latitude, (c) solar wind plasma $V_{sw}$, (d) IMF $B_0$ and $B_z$, (e) SYM-H index during May 9–13. Panels (c)–(e) and markings of the shocks (vertical dashed lines) and the wave (vertical solid line) are repeated from Figure 1 for references.

**2.3. Ionospheric Effects**

Ionospheric total electron content (TEC, representing the altitude-integrated electron number density along a path between a radio transmitting satellite and a ground receiver) measured by Swarm C satellite[13] exhibited dramatic variation during the geomagnetic storm. At the 05:00 local time (LT) sector (Figure 3a), the shock at ~17:03 UT on May 10 is found to be followed by enhanced TEC values in the equatorial region with prominent anomaly crests of >35 TEC unit (TECU) at ~15–40°N and at ~15°S geomagnetic latitudes, and a trough of ~25 TECU around the magnetic equator (TEC values are given in TECU, 1 TECU = $10^{16}$ electrons m$^{-2}$). It may be noted that 05:00 LT is quite early for a quiet-time anomaly development. Enhanced morning anomaly is found to persist during the entire main phase of the storm. After the wave impingement (at ~22:12 UT on May 10), during and well after the storm main phase, the afternoon (17:00 LT) TEC anomaly structure became stronger (Figure 3b), with a crest-to-trough TEC ratio of ~120/40 (compared to a quiet-time ratio of ~60/40), the anomaly crests shifted to higher latitudes of ~±15–45° (quiet-time crests located around ±5–15°), with an approximate latitude extent of the anomaly being ~75° (quiet-time extent of ~20°). Interestingly, the afternoon TEC values during ~15:34 UT on May 11 through ~00:20 UT on May 12 are extremely low, <20 TECU (compared to quiet-time values of ~40–50 TECU). In other words, the storm recovery is characterized by an almost "disappearing ionosphere" or an "ionospheric hole" for ~8.8 hr during the storm recovery phase.

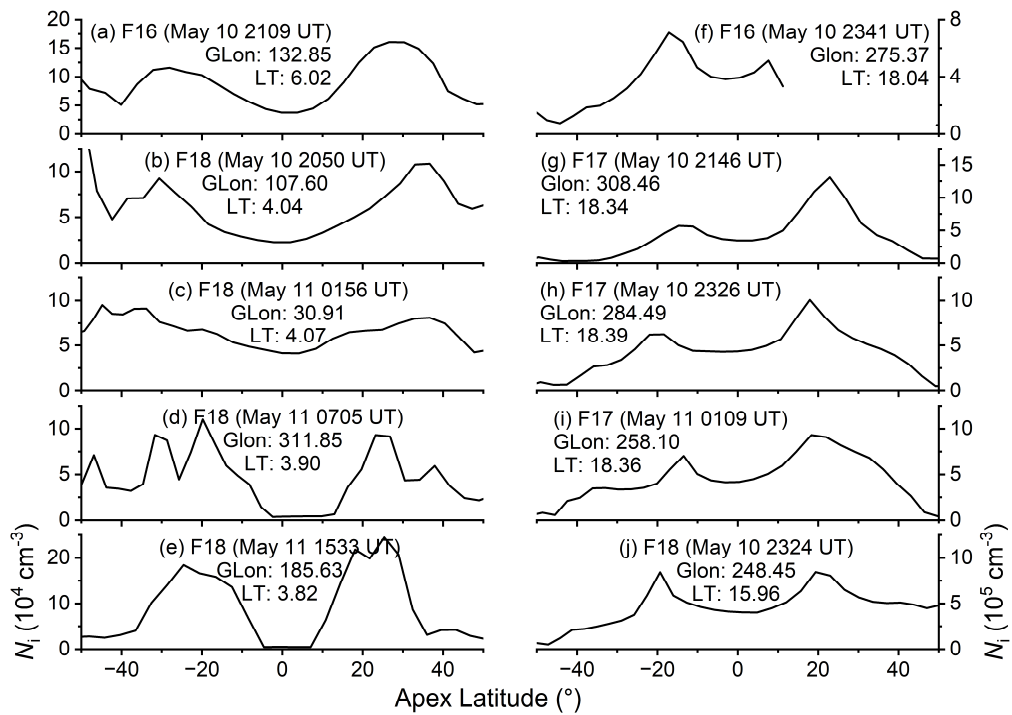

---

[13] The Swarm C satellite is one of the three-satellite Swarm constellation operated by European Space Agency (Olsen et al. 2013; Knudsen et al. 2017). The satellite was in a circular orbit at an inclination of ~87.4°. In April 2023, Swarm C orbited Earth at ~470 km altitude and has been at 05:00 and 17:00 LT. The Swarm data are obtained from https://swarm-diss.eo.esa.int.

**Figure 4. Equatorial ionization anomaly during the 2024 May storm.** Latitude variation of ionospheric ion density $N_i$ at ~850–875 km during morning (a–e) and afternoon (f–j) local times. Each panel is marked by the DMSP satellite number, date, UT, geomagnetic latitude and LT of equator crossing by the satellite.

We explored the ionospheric ion density measured by the Defense Meteorological Satellite Program (DSMP) satellites[14] at ~850–870 km in order to study the altitudinal extent of the anomaly (Figure 4). Around the period of the SYM-H peak, clear anomaly structure with two ionization crests and a trough is observed both during morning (Figures 4a–e) and afternoon (Figures 4f–j) passes of the satellites at different longitude sectors. Most interestingly, pronounced anomaly formed at ~04:00–06:00 LT, the time being too early for development of ionization anomaly under quiet geomagnetic condition. This observation is consistent with early morning TEC anomaly shown in Figure 3a.

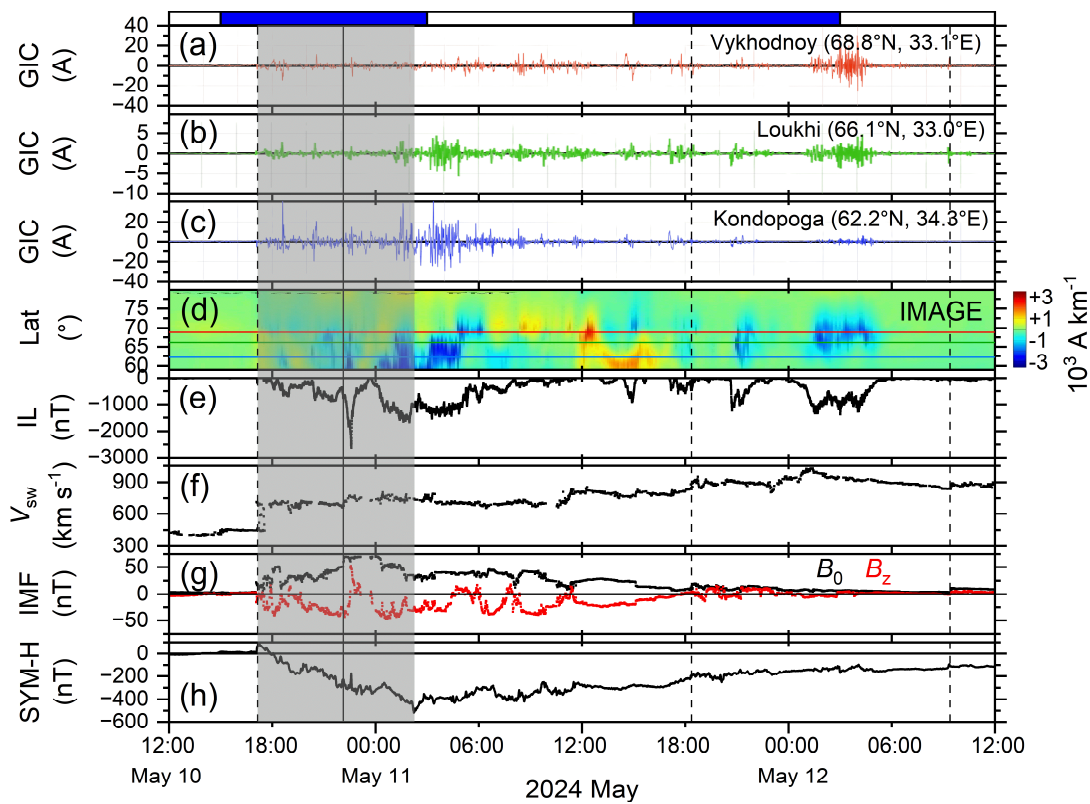

**Figure 5. Ionospheric currents during the 2024 May storm.** From top to bottom, the panels show geomagnetically induced currents (GICs) at (a) Vykhodnoy, (b) Loukhi, and (c) Kondopoga, (d) ionospheric equivalent current UT-latitude (geographic) map, (e) westward auroral electrojet index IL, (f) $V_{sw}$, (g) IMF $B_0$ and $B_z$, and (h) SYM-H during May 9–13. Local daytimes (06:00–18:00 LT) at the three

---

[14] DMSP satellite data are provided by NOAA's National Centers for Environmental Information (https://ngdc.noaa.gov/stp/satellite/dmsp/index.html).

GIC stations are marked by white horizontal bars and nighttimes (00:00–06:00 and 18:00–00:00 LT) are marked by blue bars at the top. Red, green, and blue horizontal lines in panel (d) indicate the geographic latitudes of Vykhodnoy, Loukhi, and Kondopoga, respectively. Panels (f)–(h) and markings of the shocks (vertical dashed lines) and the wave (vertical solid line) are repeated from Figure 1 for references.

**2.4. Local GIC Analysis**

The storm main phase is characterized by occurrences of multiple intense substorms, which can be identified from decreases in the IL index (Figure 5e), presenting intensification of the westward auroral electrojet currents during substorms. The IL index (Kallio et al. 2000) is based on magnetic field observations in the 16:00–03:00 UT time interval in the Fenno-Scandian region (geomagnetic latitude: 56°–76°; longitude: 96°–112°) under the 51-magnetometer International Monitor for Auroral Geomagnetic Effects (IMAGE) network[15] (Viljanen & Häkkinen, 1997). The fast forward shock at ~17:03 UT on May 10 triggered a substorm with an IL peak intensity of -692 nT at 18:56 UT (May 10), during the first-step storm main phase development. The fast forward wave at ~22:12 UT on May 10, preceded by a 2.9-hr long strong $B_s$ of 43.4 nT, led to a supersubstorm (SSS; Tsurutani et al. 2015; Hajra et al. 2016) with an IL peak of -2632 nT at 22:35 UT (May 10). This is associated with the magnetic storm second peak SYM-H. The third-step magnetic storm main phase development is associated with two intense substorms with the IL peaks of -1669 nT (at 01:58 UT on May 11), and -1531 nT (at 03:16 UT on May 11). Several intense substorms are recorded in the magnetic storm recovery phase, with the IL peaks of -911 nT (14:54 UT on May 11), -586 nT (17:35 UT on May 11), -1238 nT (20:43 UT on May 11), -1338 nT (01:39 UT on May 12), -1376 nT (03:00 UT on May 12), -1269 nT (04:05 UT on May 12), -1170 nT (22:28 UT on May 12), and -987 nT (03:08 UT on May 13). However, as the IMAGE network detects substorms occurring only in a limited time interval, many substorms occurring during the magnetic storm main and recovery phases might have not been detected.

Figure 5d shows UT-latitude map of currents flowing in the ionosphere (at an altitude of ~100 km), inferred from the ground-magnetometer observations under the IMAGE network. The magnetic storm main and recovery phases are characterized by strong eastward and westward current density (~3×10$^3$ A km$^{-1}$) with large temporal and spatial variations. Around the storm SYM-H peak, the westward currents exhibit a northward movement, followed by a southward movement of the eastward currents in the storm recovery phase.

---

[15] The IL index is obtained from the IMAGE site (http://space.fmi.fi/image/).

Figures 5a–c show the 330 kV main power line geomagnetically induced currents (GICs) measured at three stations[16] in the sub-auroral region: Kondopoga (geographic: 62.2°N, 34.3°E), Loukhi (66.1°N, 33.1°E), and Vykhodnoy (68.8°N, 33.1°E). Strong GICs are clearly triggered during the magnetic storm main and recovery phases. At Kondopoga, the fast shock at ~17:03 UT on May 10/storm onset triggered a GIC peak of ~7 A, followed by a ~40 A GIC during the first SYM-H peak. The storm main and early recovery phases are characterized by several GIC peaks of ~40 A. If we compare with the ionospheric map (Figure 5d), strong westward ionospheric current passes through the blue horizontal line (corresponding to the latitude of Kondopoga) during the times of strong GICs at Kondopoga. At Loukhi, which is northern to Kondopoga, the GIC intensity is significantly lower than at Kondopoga, ~3 A and ~5–6 A during the main and recovery phases, respectively. This result is consistent with the fact that relatively weak westward current passes through the green horizontal line (corresponding to the latitude of Loukhi). At the northern most station Vykhodnoy, GICs of ~15 A and ~30 A are recorded during the main and recovery phases, respectively. Stronger GICs in the recovery phase at Vykhodnoy corresponds to strong westward current during ~01:32–04:55 UT on May 12 passing through the red horizontal line (corresponding to the latitude of Vykhodnoy). A comparison of GICs and ionospheric equivalent currents reveal a clear association of the GICs with substorm westward currents. The GICs seem to move northward following the westward currents (Figure 5d).

---

[16] GIC data are provided by the Polar Geophysical Institute through the EURISGIC (European Risk from Geomagnetically Induced Currents) project of European Union (http://eurisgic.ru/).

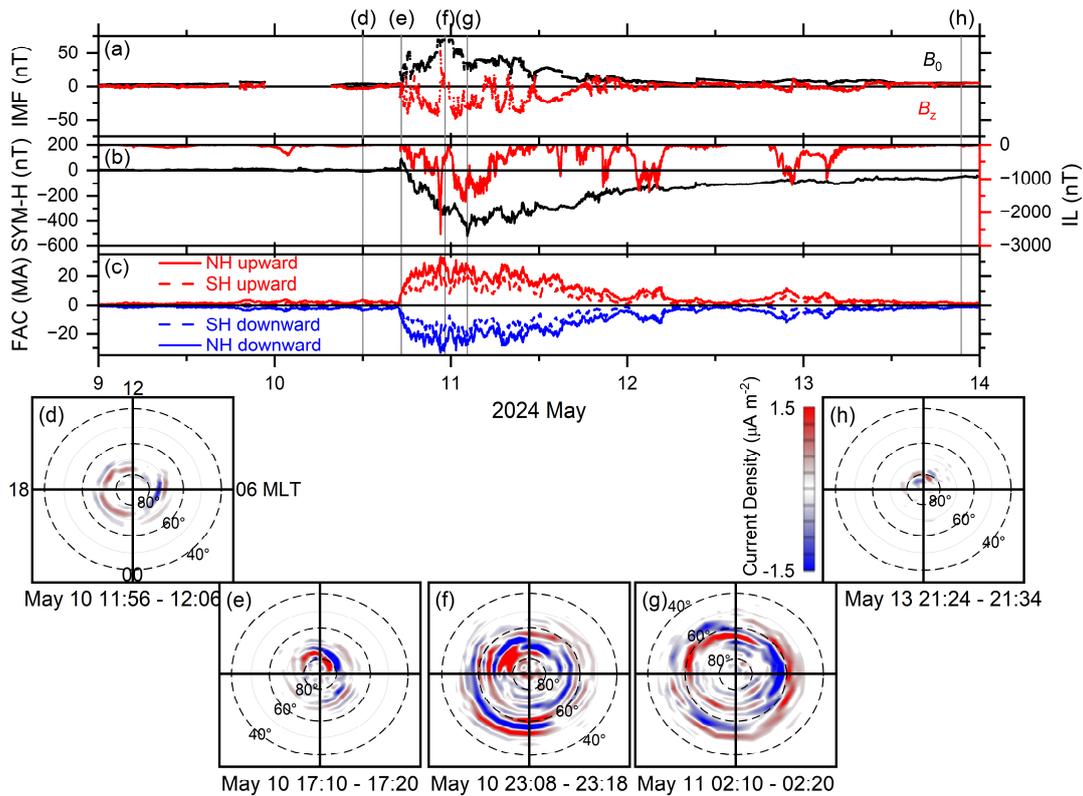

**Figure 6. Radial FACs during the 2024 May storm.** Temporal variations of (a) IMF $B_0$ and $B_z$, (b) SYM-H and IL indices, (c) radial upward (red) and downward (blue) FACs in the northern hemisphere (solid lines) and southern hemisphere (dashed lines); average radial current density for northern hemisphere plotted in AACGM and MLT coordinates during 10-minute time intervals (d) during geomagnetic quiet (e) the geomagnetic storm onset, (f) the second main phase development, (g) the third main phase SYM-H peak, (h) after the storm recovery. The 10-minute time intervals are marked by vertical gray shadings and corresponding bottom panel numbers. Red and blue in the current density panels identify upward and downward currents, respectively. Panels (a) and (b) are repeated from Figure 1 for reference.

**2.5. Ionosphere-Magnetosphere Coupling**

Figure 6c shows the temporal variations of the polar ionospheric E-region Birkeland field-aligned currents (FACs; Zmuda et al. 1966; Cummings & Dessler 1967) during the storm. The FACs are measured by the Active Magnetosphere and Planetary Electrodynamics Response Experiment[17] (AMPERE; Waters et al. 2001; Anderson et al. 2021). The upward and downward current components exhibit ~10 times increases in both northern and southern hemispheres during the storm period following the shock at ~17:03 UT on

---

[17] https://ampere.jhuapl.edu/

May 10 and southward turning of IMF compared to their pre-storm values. The peak northern hemispheric upward (downward) current of +33.4 MA (-33.26 MA) at ~22:36 UT on May 10 seems to correspond to the SSS occurring during the magnetic storm second SYM-H intensification.

Figures 6d–h show the northern hemispheric FAC maps in Altitude Adjusted Corrected GeoMagnetic (AACGM) latitude (Baker & Wing 1989) and magnetic local time (MLT) coordinate system. Figure 6d corresponds to a pre-storm/quiet period for a reference, showing only weak upward (red) Region-1 (around 70°–75° AACGM latitudes) and downward (blue) Region-2 (65°–70° latitudes) currents during noon to pre-midnight sector, and downward Region-1 and upward Region-2 currents around 06:00 MLT. Figure 6e corresponds to the SI$^+$/the storm main phase onset. This is characterized by stronger Region-1 (upward) and Region-2 (downward) currents in the dayside, associated with shock compression of (dayside) magnetosphere. Figures 6f and 6g correspond to the magnetic storm second and third SYM-H peaks, respectively. The intense substorm-related DP1 (disturbance polar) currents can be observed around the 00:00 MLT sector in a large region extending from ~50° to ~60° latitudes. In addition to this, even stronger Region-1 currents extending up to ~80° latitude, and Region-2 currents extending up to ~50° latitude, are observed in almost all local time sectors. This global-scale current system, associated with fluctuations in the magnetospheric plasma convection under strong sheath $B_s$, is called the DP2 current (Nishida 1968). After the storm recovery (Figure 6h), the current systems almost disappeared, as expected.

### 3. Summary and Discussion

We discuss below the major findings of this study on the 2024 May superstorm.

1. We identified the interplanetary causes of the superstorm (three-step main phase with SYM-H peaks of -183, -354, and -518 nT) as an interplanetary fast forward shock ($M_{ms}$ ~7.2) and a fast magnetosonic wave ($M_{ms}$ ~0.6) compressing the interplanetary plasmas and magnetic fields, leading to extremely high magnetic field magnitude of ~71 nT. The resulting interplanetary sheath was characterized by multiple strong southward IMF component $B_s$ and motional electric field $VB_s$. The three SYM-H peaks are found to correspond to $B_s$ ($VB_s$) peak values of ~40, ~43 and ~48 nT (~29, ~31, and ~35 mV m$^{-1}$) continuing for ~1.6, ~3.4, and ~4.7 hr, respectively. For a group of Dst <-280 nT storms, Echer et al. (2008) found a common interplanetary criteria: $B_s$ >20 nT ($VB_s$ >10 mV m$^{-1}$) for >3 hr. For the exceptionally intense superstorm studied here, the $B_s$ and $VB_s$ values are significantly higher than the suggested threshold values (for weaker storms), and that multiple $B_s$ components seem to lead to a greater impact. A superstorm of comparable size occurred on 2003 November 20 (SYM-H peak = -490 nT) that resulted from combined impacts of an interplanetary sheath followed by an MC (Gopalswamy et al. 2005; Echer et al. 2008). While suitable interplanetary data were not available for the 1989 March storm with a SYM-H intensity of -720 nT (strongest in the space age), it was inferred to be caused by a compound interplanetary structure of several interplanetary sheaths and a MC (see Lakhina & Tsurutani 2016; Boteler 2019; Tsurutani et al. 2024). Echer et al. (2008) studied 11 superstorms with Dst ≤ -250 nT (occurring during solar cycle 23) to conclude that ~1/3 of them were caused by interplanetary sheaths, 1/3 by MCs, and 1/3 by a combination of sheath and MC fields. More recently, Meng et al. (2019) prepared a list of all superstorms (with Dst ≤ -250 nT) occurring during 1957–2018. Their analysis suggested that "out of 19 superstorms with available concurrent solar wind data, 20% of the superstorms are caused solely by the sheath antisunward of an ICME; 10% are caused by the solar wind associated with a preceding ICME and the sheath antisunward of the present ICME, that is, compound ICMEs; 45% are caused by the sheath antisunward and the magnetic cloud of an ICME; 5% are caused solely by the magnetic cloud of an ICME." However, none of these studies reported a superstorm caused by a shock-sheath and a wave-sheath interactions, like the present one.

2. The Forbush decreases in the CR counts recorded during the storm main phase ranged from ~11% at Oulu (Finland, effective vertical cutoff rigidity ~0.8 GV) to ~17% at Dome C (Antarctica, rigidity <0.01 GV). At Dome C, the decrease is prominently a two-step event. Variation in the decreases (from one station to another) is related to the neutron monitor type, the cutoff rigidity, and altitude of the monitor. However, these decreases are in the range of the large Forbush decreases, i.e. ~10–25% (e.g. Cane 2000). The largest decrease on record is ~35%, recorded at the South Pole, Antarctica (geomagnetic latitude: 80.7°S, longitude: 107.3°E, altitude above sea level: 2820 m, cutoff rigidity: 0.1 GV) on 1972 August 5 during a geomagnetic storm with a Dst peak of -107 nT (Pommerantz & Duggal 1973). CR Forbush decreases are attributed mainly to fast ICMEs pushing the CR particles away from the Earth (Simpson 1954; Cane 2000, and references therein). The interplanetary sheath following the shock and the ICME "ejecta"/MC are suggested to be responsible for two-step CR decreases (Cane et al. 1994; Janvier et al. 2021).

3. The storm main phase was characterized by significant losses of ~76 keV to ~1.5 MeV electrons (with no significant impact on the ~2.0–2.9 MeV electrons) in the geosynchronous orbit following impingement of the $M_{ms}$ = 7.2 fast forward shock on the magnetopause. This can be explained due to a "magnetopause shadowing" effect (West et al. 1972, 1981). The shock compresses the dayside magnetospheric outer zone magnetic fields, making them blunter than a dipole configuration. Energetic electrons gradient drifting from the midnight to the morning sector will drift toward the magnetopause boundary and be lost to the magnetosheath. This magnetopause shadowing effect may lead to electron losses on open drift paths. However, this shock did not exhibit any impact on ≥2.0 MeV electrons probably owing to low flux levels of these electrons during the pre-shock interval. In addition, other two shocks and a steepened wave exhibited no apparent impacts on the ~76 keV to ~2.9 MeV electrons in the geosynchronous orbit. This result seems to be surprising. Hajra and Tsurutani (2018b) reported significant decreases of the geosynchronous orbit >0.8 and >2.0 MeV electrons following an $M_{ms}$ = 2.9 shock. Hajra et al. (2020) reported that the entire outer belt (including the geosynchronous orbit) was depleted of ~1.0–4.5 MeV electrons following an $M_{ms}$ ~6.7 shock. Clearly, more studies are required to understand the magnetospheric electron losses due to shock compression, considering possible contributions of the shock characteristic parameters. More recently, Hajra et al. (2024) reported depletion of the outer belt ~2.0–2.9 MeV electrons in association with the ram pressure compression of the magnetosphere during corotating interaction regions (CIRs). CIRs form between high-speed solar wind streams emanated from solar coronal holes and slow solar winds, and are characterized by amplified Alfvén waves (Smith & Wolfe 1976; Tsurutani et al. 2006a). As both shock and CIR compress the magnetosphere, electron loss mechanisms during both events might be identical. Hajra et al. (2024) suggested that, in addition to the magnetopause shadowing effect, the magnetospheric compression by enhanced ram pressure can also excite electromagnetic ion cyclotron (EMIC) waves (Thorne & Kennel 1971; Horne & Thorne 1998) in the dayside magnetosphere, and the cyclotron resonant interaction of the relativistic electrons with the EMIC waves is another possible loss mechanism for these electrons to the ionosphere (Remya et al. 2015; Tsurutani et al. 2016). This mechanism should also be verified to better understand varying impacts of the interplanetary shocks (and their efficiency) on the magnetospheric electrons.

4. During the magnetic storm main phase, the magnetosphere is inflated in size by the formation of the ring current. As the ring current particles are lost in the magnetic storm recovery phase, the magnetosphere deflates and the magnetic field lines threading the ring current move inward and the magnetic field intensifies. This will cause an apparent "radial diffusion" and also a betatron acceleration of the very high energy particles which have remained trapped. This mechanism is consistent with gradual increases in the ~0.9–2.9 MeV electron fluxes in the recovery phase. Another possibility is the wave-particle interaction leading to the electron acceleration. The sporadic ~76–132 keV electron injections during the storm recovery (observed in this work) can lead to whistler-mode chorus wave generation owing to the temperature anisotropy of the electrons (Kennel & Petschek 1966; Tsurutani & Smith 1977; Meredith et al. 2001). Resonant cyclotron interactions of the ~100 keV electrons with the chorus waves can effectively accelerate the electrons to ~MeV electrons (e.g. Inan et al. 1978; Horne & Thorne 1998; Tsurutani et al. 2006b; Summers et al. 2007; Reeves et al. 2013; Boyd et al. 2014).

5. Triggering of the early morning (~04:00–05:00 LT) ionospheric ionization anomaly in the (magnetic) equatorial region following the interplanetary shock is an important result of this work. As confirmed by Swarm C satellite 05:00 LT and 17:00 LT passes, the dayside ionosphere anomaly was amplified, in terms of enhancements of the anomaly crest plasma density (by a factor of ~2) and expansion of the latitudinal extent (by ~3.75 factor). DMSP satellites confirmed high-altitude anomaly structure at ~850–875 km, beyond the quiet-time ionospheric F2 region or uplifting of the F2 layer. These observations are consistent with creation of a "dayside superfountain effect" due to strong prompt penetration electric field reaching the equatorial F2 region ionosphere (Tsurutani et al. 2004, 2008; Mannucci et al. 2005). Because of the ***E×B*** convection of the plasma, the ionospheric anomalies reach higher magnetic latitudes. During the 2003 October 30–31 "Halloween" superstorm, the anomalies reached ~±30° magnetic latitudes, while for the present event they reached ±45° magnetic latitudes (instead of the usual ±10° magnetic latitudes during quiet times). When the plasma is lifted to higher altitudes, the recombination rate there is much lower than at lower altitudes. Therefore, the recombination of ions with thermal electrons back into neutrals is substantially decreased. Meanwhile solar photons are creating new ionospheric plasma at lower altitudes replacing the plasma that has been uplifted, increasing the overall TEC. Another interesting ionospheric impact is creation of an afternoon "ionospheric hole" (exceptionally low TEC) during the storm recovery phase. According to Fejer and Scherliess (1995), strong westward electric field is created in the equatorial region due to storm-time (disturbance) dynamo effect of global thermospheric wind circulation generated by polar region Joule heating (due to solar energy injection and particle precipitation) during the storm recovery phase. This westward electric field may restrict the anomaly formation, leading to low plasma density in the equatorial region.

6. The geomagnetic storm main and recovery phases are characterized by multiple intense sub-auroral region substorms, including a SSS in the main phase. Hajra et al. (2016) reported a lack of statistical association of the SSS intensity with the geomagnetic storm intensity. However, Tsurutani & Hajra (2023) reported several SSS events occurring simultaneously with the concurrent magnetic storms. Present observations support the idea of complex substorm–storm relationships. Based on the IMAGE ground-magnetometer observations the inferred equivalent ionospheric currents are found to exhibit large temporal and spatial variations, consistent with a northward movement of GICs during the storm main and recovery phases. Peak GICs of ~30-40 A are recorded during the storm. The strongest GIC intensity of 57 A at Mäntsälä (geographic: 60.6°N, 25.2°E) was associated with a SSS (SML peak of -3548 nT) occurring during the 2003 October 29–30 "Halloween" superstorm (SYM-H peak of -390 nT) (Tsurutani & Hajra 2021). In the present work, the SSS is found not to be associated with the strongest GICs, confirming that there exists no linear relationship of a GIC intensity with a SSS or a magnetic storm.

7. The storm period is characterized by strong Region-1 and Region-2 FACs, with ~10 times increases during the storm main phase compared to their pre-storm values. During the peak SYM-H developments of the storm, the Region-1 exhibited large poleward expansion up to ~80° latitude, and Region-2 down to ~50° latitude. These results are indicative to large-scale and strong magnetic convection associated with southward IMFs during this giant geomagnetic storm. In addition to the substorm-related midnight sector auroral DP1 currents, intensification of global-scale DP2 currents extending from ~80° to ~40° geomagnetic latitude is consistent with the worldwide auroral displays down to unusually low latitudes. This also corroborates with recent suggestion (Tsurutani & Hajra 2023) of global-scale magnetospheric/ionospheric energy dissipation (Nishida 1968) during intense substorm and magnetic storm activity.

## Acknowledgments

The work of RH is funded by the "Hundred Talents Program" of the Chinese Academy of Sciences (CAS), and the Excellent Young Scientists Fund Program (Overseas) of the National Natural Science Foundation of China (NSFC). We gratefully acknowledge the following data sources: NASA's Coordinated Data Analysis Web (CDAWeb: https://cdaweb.gsfc.nasa.gov/istp_public/) for the solar wind plasma and IMF measured by the WIND spacecraft; OMNIWeb Plus database provided by NASA (https://omniweb.gsfc.nasa.gov/) for the solar wind plasma and IMF data shifted to the Earth's bow shock nose; the Solar Dynamic Observatory (https://sdo.gsfc.nasa.gov/) for the solar image; the World Data Center for Geomagnetism, Kyoto, Japan (https://wdc.kugi.kyoto-u.ac.jp/) for the geomagnetic Dst index; the Cosmic Ray Station of the University of Oulu/Sodankyla Geophysical Observatory and the French-Italian Concordia Station (IPEV program n903 and PNRA Project LTCPAA PNRA14_00091) (https://cosmicrays.oulu.fi/) for the CR data; the IMAGE site (http://space.fmi.fi/image/) for the IL index and ionospheric equivalent currents; GOES-18 (https://www.ngdc.noaa.gov/stp/satellite/goes-r.html) for the outer radiation belt proton and electron fluxes; the European Space Agency (https://swarm-diss.eo.esa.int) for the Swarm C satellite data; NOAA's NCEI site (https://ngdc.noaa.gov/stp/satellite/dmsp/index.html) for the DMSP satellite data; the Polar Geophysical Institute for the GIC data through the EURISGIC project (http://eurisgic.ru/), and AMPERE (https://ampere.jhuapl.edu/) for the polar region radial FACs. We would like to thank the reviewer for extremely valuable suggestions that substantially improved the manuscript.

## Appendix A
## Active Region AR3664, Solar Flares, and CMEs

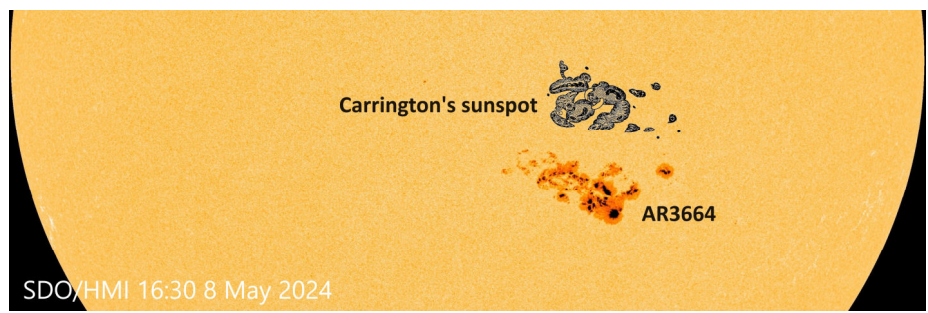

**Figure A1: Solar active region AR3664 in comparison to the "Carrington's sunspot".** A close-up of the solar image taken at 16:30 UT on 2024 May 8 showing AR3664 along with the Carrington's sketch (to scale) of sunspot observed on 1859 September 1. Image modified from: https://spaceweather.com/.

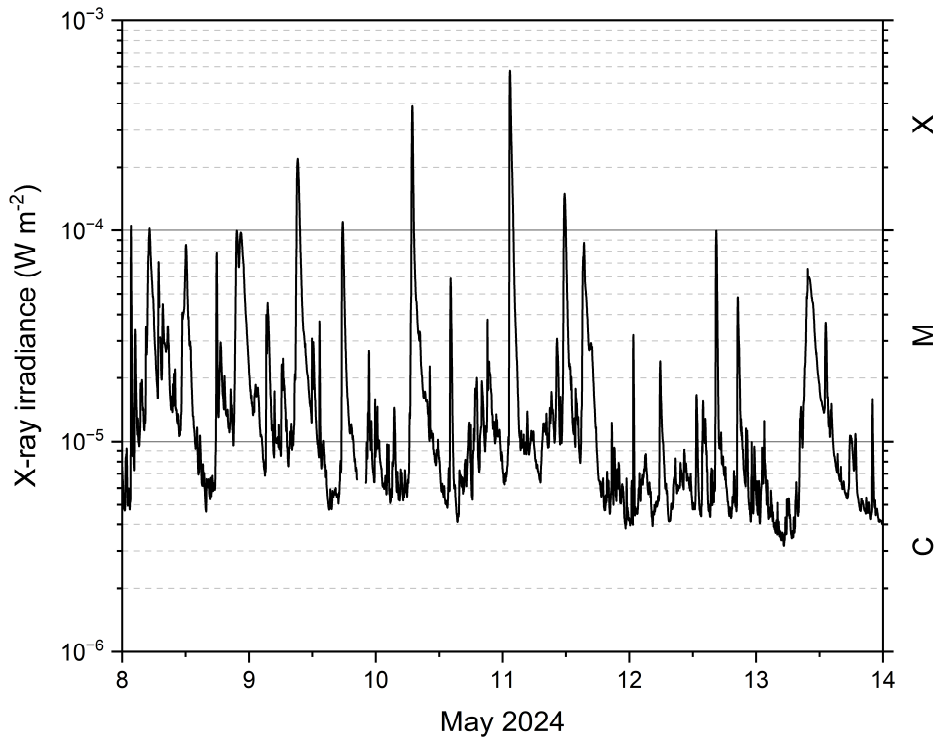

**Figure A2: GOES-18 X-ray irradiance (wavelength range of 1–8 Å) during 2024 May 8–13**. The classes of X-ray flares are indicated on the right.

The solar active region AR3664 (Figure A1) was giant in size, comparable to the "Carrington" sunspot (Carrington, 1859), and had an unstable "β-γ-δ" magnetic field (a sunspot group with a bipolar sunspot group (β) but complex enough so that no line can be drawn between spots of opposite polarity (γ), but contains one (or more) sunspot(s) with the opposite polarity umbrae in a single penumbra (δ)) that harboured energy for several X-class solar flares during May 8–13 (Figure A2): X1.1 (at 01:41 UT on May 8), X1.0 (05:09 UT on May 8), X2.2 (09:14 UT on May 9), X1.1 (17:44 UT on May 9), X3.9 (06:54 UT on May 10), X5.8 (01:23 UT on May 11), and X1.5 (11:44 UT on May 11). Multiple CMEs erupted from AR3664 in association of those solar flares. The following ICMEs caused multiple fast forward shocks and a wave followed by interplanetary sheaths, shown in Figure 1.